# Ethical AI prompt recommendations in large language models using collaborative filtering

Jordan Nelson[1], Almas Baimagambetov[1], Konstantinos Avgerinakis[2] and Nikolaos Polatidis[1],

[1]School of architecture Technology and Engineering, University of Brighton, BN2 4GJ, U.K
[2]Catalink Limited, Nicosia, Cyprus

**Abstract**: As large language models (LLMs) shape AI development, ensuring ethical prompt recommendations is crucial. LLMs offer innovation but risk bias, fairness issues, and accountability concerns. Traditional oversight methods struggle with scalability, necessitating dynamic solutions. This paper proposes using collaborative filtering, a technique from recommendation systems, to enhance ethical prompt selection. By leveraging user interactions, it promotes ethical guidelines while reducing bias. Contributions include a synthetic dataset for prompt recommendations and the application of collaborative filtering. The work also tackles challenges in ethical AI, such as bias mitigation, transparency, and preventing unethical prompt engineering.

## 1. Introduction

As large language models (LLMs) have been playing one of the central roles in AI and Machine Learning (ML) development for years, ensuring that these generations adhere to ethical standards becomes paramount. Because of this, prompt recommendations for generating models and algorithms in this domain becomes increasingly crucial in addressing ethical adherence as a pre-emptive measure before even reaching the generative stage of an LLMs output. While LLMs have the potential to accelerate the creation of complex AI solutions, they also introduce risks related to fairness, bias, and accountability. Without proper oversight, LLM-generated models may inadvertently perpetuate harmful biases, optimise for unintended objectives, or reflect unethical practices. These risks are particularly concerning in high-stakes applications such as healthcare, finance, and criminal justice, where biased or unethical AI systems can have real-world consequences [33].

Traditional methods of ensuring ethical AI, such as rule-based filtering [1] and human moderation, often struggle with scalability and adaptability, making it challenging to enforce evolving ethical standards. Rule-based approaches rely on predefined constraints that may not account for nuanced ethical dilemmas or emerging AI risks, while human moderation is labour-intensive and unfeasible at scale. As AI research advances and new ethical challenges arise, static oversight mechanisms [2] become increasingly insufficient. This highlights the need for more dynamic, data-driven approaches to ensure responsible model generation, allowing AI systems to adapt to evolving best practices while maintaining a high level of ethical scrutiny. One such approach is collaborative filtering, a technique commonly used in recommendation systems to personalise content based on user interactions, preferences, and feedback. By applying collaborative filtering to prompt recommendations for AI model generation, LLMs can learn from user behaviour, promoting prompts that align with ethical guidelines and discouraging those that could lead to biased or unsafe AI models. This data-driven approach enables the recommender system to evolve over time, adapting to shifts in ethical considerations and reinforcing best practices. Additionally, by incorporating mechanisms for semantic similarity and feedback loops, collaborative filtering can help identify emerging trends in AI ethics, guiding users toward prompts that prioritise fairness, transparency, and accountability.

This paper proposed a collaborative filtering methodology that can be adapted to improve the ethical recommendations of prompts for generating LLM-based AI models. Therefore, the contributions of this paper are as follows:

- A synthetic dataset for prompt recommendations in LLMs.
- An application of collaborative filtering to the dataset.

Given the complexities of ethical AI design, it is crucial that recommendation systems do not reinforce existing biases, unfairly prioritise certain modelling approaches, or overlook risks like adversarial manipulation. Therefore, we also address key challenges, including bias mitigation, transparency, and the prevention of unethical prompt engineering.

The remaining paper is structured as follows: section 2 delivers the related work, section 3 explains the dataset and the proposed methodology, section 4 presents and discusses the results, and section 5 contains the conclusions.

## 2. Related work

LLMs have recently found applications across numerous domains, ranging from general-purpose chatbots [3, 4] capable of engaging in conversation and answering questions to specialised assistants designed for various tasks [5]. Their rapid rise in social popularity led to widespread adoption in both industry and academic research, further driving innovation in the field.

One such use is the generation of machine learning code and models. Recent studies into code generation by state-of-the-art (SOTA) LLMs such as ChatGPT [3] are limited due to the novelty of these technologies, and the full extent of their capabilities is still being explored and expanded upon. Consequently, the available research on this topic remains scarce. Some publications [6, 7] have identified several flaws and security issues in the code generated by ChatGPT. For instance, out of 21 reported use-cases, only five were initially secure, with an additional seven becoming more secure only after explicit instructions were provided by the user. Other studies have revealed that synthetic code generated by ChatGPT exhibited vulnerabilities in over a third of their use-cases, with some studies reporting vulnerabilities in approximately 12% of cases [8, 9]. A study [10] examining machine learning code produced by LLMs found that although 58% of the code generated was correct, we cannot ignore a large portion amounting to 42% contained errors. This highlights significant room for improvement in the reliability of LLM-generated code. Other research [11] has explored the integration of existing tools, such as AutoML, to mitigate these shortcomings. While LLM-generated code can be effective for straightforward tasks, it often falls short when tasked with developing more complex, accurate, and bug-free machine learning algorithms.

A separate study [12] investigating the security of LLM-based code generation uncovered notable vulnerabilities, including issues with the handling of security-sensitive values, susceptibility to SQL injection, and the presence of hard-coded values that would need adjustments before the code could be considered suitable for deployment. Meanwhile, other research [13, 14] has gone beyond analysing the code itself, focusing instead on the evaluation conditions and metrics used to assess the performance of LLM-generated code. These studies suggest that even the benchmarks and methodologies for evaluating such code are being critically examined, raising questions about how to define "good code". Additionally, a broader analysis [15] of 2,033 programming tasks revealed frequent problems in ChatGPT-generated code, such as compilation errors, incorrect outputs, maintainability issues, and inefficiencies, further highlighting the areas needing improvement in AI-driven code generation.

Research into the ethical implications of LLMs has increasingly focused on accountability, hallucinations, bias, and data privacy. Hallucinations where models generate false or misleading information pose significant risks, particularly in high-stakes domains like medicine. Ensuring accountability is challenging due to the black-box nature of LLMs, making it difficult to trace errors. Researchers have explored mitigation strategies such as improved training datasets, reinforcement learning, and transparency measures [16]. In medicine, inaccurate AI-generated insights raise concerns about bias, informed consent, and over-reliance on automated recommendations [17]. Bias in LLMs stems from training data, leading to discriminatory or harmful outputs. Studies have examined debiasing techniques, adversarial training, and diverse dataset curation to reduce these risks [18]. Additionally, LLMs can generate misinformation, which may be exploited maliciously. Ethical safeguards such as content moderation, model transparency, and stricter guidelines have been proposed to address these concerns [19].

Data privacy and security are also critical, particularly in fields like medicine and education. LLMs process vast amounts of sensitive data, raising concerns about compliance with regulations like GDPR and HSE. In medical applications, improper data handling risks patient confidentiality, while in education, AI-driven tools must balance personalisation with data protection [20]. Researchers have explored solutions such as federated learning and differential privacy to enhance security and prevent misuse [21, 22].

Recommender systems are essential for personalising content [23, 24], and with the rise of LLMs are being applied in many ways. By analysing user prompts and coding patterns, these systems can suggest relevant snippets or algorithms. However, integrating LLM-generated code introduces challenges like code quality, security risks, and ethical concerns, such as bias and intellectual property. Ensuring the accuracy and responsibility of these recommendations is crucial for their effective use.

One study [25] explored the integration of an LLM into the recommender process, highlighting several benefits but also acknowledging that it faced challenges like those of LLMs in general, such as hallucinations, safety issues, and biases. In a similar vein, research on prompt guidance (PG) [26] suggests that providing PG in LLM-based systems can significantly enhance user interactions, particularly in domains where accuracy is critical. PG helps users express their preferences more clearly and allows the system to refine its responses, resulting in more accurate and satisfying recommendations [27]. When applied to LLM-based code generation, PG could offer similar advantages, helping users, especially non-technical ones, overcome the complexities of prompt engineering. By guiding them to create more precise coding prompts, PG can enable the generation of better code snippets and debugging suggestions. Furthermore, PG is especially beneficial in high-stakes environments, such as generating code for crucial applications, where ensuring accuracy and giving users more control is essential [28].

## 3. Proposed methodology

To solve the problem of prompt recommendation we utilise a two-agent system. As shown in figure 1, the first agent is an LLM such as ChatGPT and the second agent is a simple but efficient collaborative filtering algorithm. Initially, the LLM passes each submitted prompt to the collaborative filtering algorithm which then calculates a similarity with other prompts found in its dataset and the prompts with the highest similarity are recommended as follow up prompts in the LLM. The main idea is to help the user of an LLM to choose relevant prompts instead of typing prompts, thus making the process faster and more efficient, while ensuring the prompts are recommended based on similar users' preferences.

Collaborative filtering is a recommendation technique that identifies similarities between users or items based on historical interaction data, such as ratings or preferences. It does not evaluate content directly but instead uses patterns in how users interact with items to make predictions. In our context, each prompt is rated in relation to other prompts by simulated users, and these ratings form the basis for computing similarity. The system learns to recommend prompts that users with similar interaction histories found relevant. This approach enables ethical recommendations to be recommended from collective user feedback.

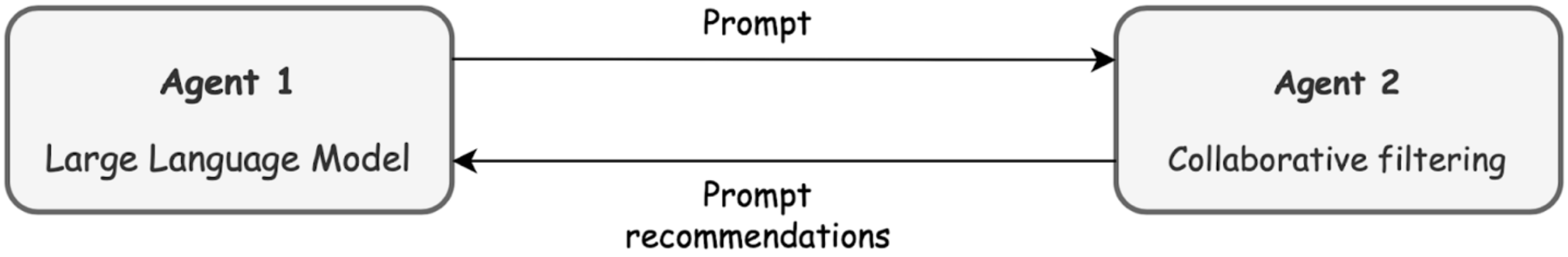

**Figure 1.** A multi-agent prompt recommender system

The dataset used is represented as follows: Prompt, Prompt, Rating. This was inspired by the well-known MovieLens dataset where users rate movies in the form of userid, movieid, rating [29]. In this case instead of users and movies we have prompts and for each prompt there is a 1 to 5 rating with other prompts. To the best of our knowledge there are no such datasets available in the literature, thus we utilised ChatGPT 4o to create a synthetic dataset with 3612 entries with values from 1 to 5 which can be found at: https://www.kaggle.com/datasets/jordanln/prompt-recommendations-for-llm

While our dataset offers a synthetic environment for evaluating ethical prompt recommendations, it does not capture the full variability of real-world user input. In practical settings, user-submitted prompts can be highly diverse, context-dependent, and domain-specific, often reflecting ambiguous or nuanced ethical considerations. Incorporating actual prompts from users interacting with LLM-based systems from domains such as education or healthcare could potentially improve the dataset and the accuracy of the recommender system. Moreover, in table 1 we show some of the entries of the dataset. Finally, to determine the similarity between two prompts, our system uses the Pearson correlation coefficient, a standard metric in collaborative filtering. This metric evaluates how similarly users have rated two different prompts, helping the system recommend prompts that have received similar feedback from users with comparable preferences.

Let a and b be two prompts for which we want to compute the similarity. We define:

- $r_{u,a}$: The rating given by user u to prompt a
- $r_{u,b}$: The rating given by user uu to prompt bb
- $\bar{r}a$: The average rating given to prompt aa across all users who rated it
- $\bar{r}b$: The average rating given to prompt bb across all users who rated it
- U: The set of users who rated both prompts a and b

The Pearson correlation coefficient sim(a,b) is then calculated as follows:

$$\text{Sim}\,(\text{a}, \text{b}) = \frac{\sum_{p\in P}(r_{a,p} - \overline{r_a})(r_{b,p} - \overline{r_b})}{\sqrt{\sum_{p\in P}(r_{a,p} - \overline{r_a})^2}\sqrt{\sum_{p\in P}(r_{b,p} - \overline{r_b})^2}} \quad (1)$$

Additionally, it should be noted that Pearson correlation as shown in the equation above allows the system to identify how similarly two prompts are rated by the same users not whether they are rated highly overall. For example, two prompts might both have moderate average ratings but tend to be rated similarly by users who like ethical models. This pattern of agreement is what Pearson correlation captures. Besides that, to ensure the system remains adaptable over time, such recommendation algorithms support dynamic dataset evolution. This means new prompts can be added as users interact with the system. This helps the system to handle new inputs while continuously improving its recommendation quality.

Finally, from an infrastructure perspective, the prompt-rating dataset can be implemented in either a centralized or domain-specific manner, depending on the deployment scenario. A centralized, public dataset could facilitate broader coverage with shared ethical standards, and collective learning across a wide user base. However, for organizations with specific regulatory or domain constraints such as education or healthcare, specialized datasets can offer more targeted and context-sensitive recommendations while reducing the overall overhead. Both approaches are suitable and can be implemented using collaborative filtering.

| Prompt | Prompt | Rating |
|---|---|---|
| Design a recommendation system that avoids bias. | Design an image recognition system minimizing gender bias. | 1.25 |
| Design a recommendation system that avoids bias. | Create a predictive model for environmental impact analysis. | 3.09 |
| Design a recommendation system that avoids bias. | Generate an AI-based tutoring system ensuring fairness. | 2.85 |
| Design a recommendation system that avoids bias. | Develop an ML model for equitable healthcare resource allocation. | 2.52 |
| Design a recommendation system that avoids bias. | Generate an AI-based tutoring system ensuring fairness. | 1.25 |
| Build a sentiment analysis model with privacy constraints. | Design an image recognition system minimizing gender bias. | 1.78 |
| Build a sentiment analysis model with privacy constraints. | Design a recommendation system that avoids bias. | 2.24 |
| Build a sentiment analysis model with privacy constraints. | Generate an AI-based tutoring system ensuring fairness. | 4.47 |
| Build a sentiment analysis model with privacy constraints. | Build a chatbot that supports diverse language inclusivity. | 3.35 |
| Build a sentiment analysis model with privacy constraints. | Develop an ML model for equitable healthcare resource allocation. | 4.33 |
| Develop an ML model for equitable healthcare resource allocation. | Design a recommendation system that avoids bias. | 2.2 |
| Develop an ML model for equitable healthcare resource allocation. | Create a fraud detection algorithm preventing racial profiling. | 1.35 |
| Develop an ML model for equitable healthcare resource allocation. | Generate a privacy-preserving NLP model for email filtering. | 1.36 |
| Develop an ML model for equitable healthcare resource allocation. | Generate an AI-based tutoring system ensuring fairness. | 4.36 |
| Develop an ML model for equitable healthcare resource allocation. | Design an image recognition system minimizing gender bias. | 4.97 |

**Table 1.** Snapshot of ratings from the dataset

## 4. Experimental evaluation

The experimental evaluation took place on a computer running the linux operating system along with the Python programming language version 3.14 and the surprise library version 1.1.4. In section 4.1 we define the evaluation metrics, in 4.2 we present the results and in 4.3 we discuss the limitations.

The goal of the experimental evaluation is to assess how accurately the proposed collaborative filtering system recommends relevant prompts based on user interaction data. Specifically, we measure how well the algorithm predicts ratings for prompt-prompt pairs and identify top-rated prompts that align with a given input prompt.

The evaluation uses the synthetic dataset of 3,612 prompt-prompt-rating entries, where each entry consists of:

1. A base prompt.
2. A related prompt.
3. A synthetic rating (on a scale of 1 to 5) indicating their ethical or contextual relevance.

The collaborative filtering model is trained using this dataset to learn patterns of similarity between prompts. During testing, the system takes a given prompt as input and produces a ranked list of recommended prompts as output. These predictions are then compared against the known ratings in the dataset.

Additionally, during testing, each prompt in the test set is treated as a query, and the system is asked to return the top 10 recommended prompts for it. This simulates a realistic user scenario where a user enters a single prompt and expects a short list of relevant suggestions. These 10 recommended prompts are then compared against the actual known ratings in the dataset to determine how many of them meet the relevance threshold (e.g., 3.0 or 3.5).

### 4.1 Evaluation metrics

For the evaluation we have used well known evaluation metrics that have been widely used in recommender systems [30-32]. The metrics are the Mean Absolute Error (MAE), Root Mean Squared Error (RMSE), Precision, Recall and F1. MAE measures the average magnitude of the absolute differences between predicted ratings and actual ratings. It indicates how far the predictions deviate from the true ratings and is defined in equation 2. In this context, n is the total number of ratings in the dataset. The value p refers to the actual rating that a prompt received, and r is the rating that the system predicted for that same prompt. These predicted ratings are generated by the collaborative filtering algorithm based on patterns it learns from the training data. RMSE measures the square root of the mean squared differences between predicted and actual ratings. It penalises larger errors more than MAE, making it sensitive to outliers and is defined in equation 3 and again as previously $n$ is the total number of ratings, $p$ is an actual rating, and $r$ is a predicted rating for a prompt $i$. Precision is used to measure the proportion of recommended items that are relevant to the user. It answers the question: "Of all recommended items, how many were actually liked?" and is defined in equation 4 where TP stands for True Positive meaning the correct recommendation was made and FP stands for False Positive meaning the incorrect recommendation was made. Recall measures the proportion of relevant items that were successfully recommended. It answers the question: "Out of all relevant items, how many were recommended?" and is defined in equation 5 where TP stands for True Positive and FN stands for False Negative meaning a correct recommendation was not made. Finally, F1 which is defined in equation 6 balances precision and recall into a single metric using their harmonic mean. It is useful when both false positives and false negatives are important.

$$MAE = \frac{1}{n}\sum_{i=1}^{n}|p_i - r_i| \quad (2)$$

$$RMSE = \sqrt{\frac{1}{n}\sum_{i=1}^{n}(p_i - r_i)^2} \quad (3)$$

$$Precision = \frac{TP}{TP + FP} \quad (4)$$

$$Recall = \frac{TP}{TP + FN} \quad (5)$$

$$F1 = \frac{2xPrecisionxRecall}{Precision + Recall} \quad (6)$$

### 4.2 Results

The results are based on a request of 10 recommendations using a threshold of 3.0 out of 5 and a threshold of 3.5 out of 5. Both tests have used a 10-fold cross validation method. In this experiment, a threshold is used to determine which prompt recommendations are considered relevant. Specifically, we test two threshold values: 3.0 and 3.5 out of a maximum rating of 5.0. This means that only recommendations with a predicted rating equal to or above the threshold are counted as correct. For example, if the threshold is 3.0, then any prompt with a rating of 3.0 or higher is considered a good recommendation.

| Threshold | MAE | RMSE | Precision | Recall | F1 |
|---|---|---|---|---|---|
| 3 | 1.0225 | 1.1834 | 1.0000 | 0.9960 | 0.9980 |
| 3.5 | 1.0223 | 1.1840 | 0.6000 | 0.0540 | 0.0990 |

**Table 2.** Snapshot of ratings from the dataset

Lowering the threshold to 3.0/5 drastically improves precision and recall, leading to a near-perfect F1-score. However, this might also mean more recommendations overall, potentially increasing noise if users expect only high-confidence suggestions. The choice of threshold should balance precision and recall based on the intended user experience. These results indicate that reducing the threshold leads to significant improvements in recall and F1-score, but the increase in recall may also introduce more diverse recommendations. The best threshold depends on the desired trade-off between precision and recall in the recommender system.

### 4.3 Limitations

In the context of a two-agent system for ethical machine learning prompt recommendations, a key challenge arises when a user-entered prompt does not match exactly with any existing prompt in the dataset. This scenario highlights the limitations of traditional exact matching mechanisms and underscores the importance of implementing robust handling strategies. Exact matches rely on predefined entries in the dataset, which may not accommodate user-specific phrasing or entirely new prompt ideas. As a result, the system must employ alternative approaches to ensure meaningful recommendations.

One effective method is to leverage semantic similarity techniques, where the system computes the similarity between the user-entered prompt and prompts in the dataset using language models such as BERT or GPT embeddings. By transforming prompts into vector representations, the system can calculate cosine similarity to identify the most contextually relevant prompts, even when there is no direct textual overlap. This approach not only enhances the system's flexibility but also ensures that the recommendations remain contextually relevant and aligned with the user's intent. For example, a prompt about “developing privacy-preserving NLP models” could still retrieve suggestions related to federated learning or differential privacy, even if the wording differs.

Alternatively, in cases where semantic similarity fails to yield close matches, the system can resort to fallback strategies such as providing default or popular recommendations. These recommendations could be derived from the highest-rated prompts in the dataset or pre-defined ethical considerations that are universally relevant. Additionally, the system could employ a dynamic expansion mechanism, where new user prompts are added to the dataset and linked with inferred ratings. This approach allows the dataset to evolve and adapt over time, capturing a broader spectrum of user needs and fostering a more comprehensive recommendation model.

By combining semantic matching with fallback mechanisms and dynamic dataset expansion, the system can address the inherent variability in user input while maintaining the quality and relevance of recommendations. This not only enhances user satisfaction but also ensures the scalability and adaptability of the two-agent system in dynamic research and development environments.

One other limitation can be found in the exclusive use of synthetic data. Although synthetically generated prompts allow for experimentation, they may not fully represent the complexity and of real user inputs. Actual user prompts can introduce varied phrasing, implicit ethical trade-offs, or domain-specific language that the synthetic dataset does not capture. To solve this real-world prompt data, need to be collected.

One potential limitation of the proposed system is its susceptibility to manipulation through coordinated or repeated malicious input. For example, an attacker with sufficient computational resources could repeatedly submit a harmful prompt and rate it highly in relation to other prompts, thereby inflating its perceived relevance. While collaborative filtering helps personalise recommendations based on shared patterns across users, it does not inherently detect or prevent adversarial behaviour of this kind. Addressing such risks would require additional mechanisms, such as anomaly detection, adversarial training, or trust-weighted rating schemes.

## 5. Conclusions

This study has explored a recommender system for ethical prompt suggestions in large language model-assisted machine learning code generation. The findings underscore the importance of balancing precision and recall ensuring the system provides relevant, high-quality prompts while maintaining flexibility and adaptability. By leveraging collaborative filtering, semantic similarity techniques, and dynamic dataset expansion, the approach enhances the recommendation process, addressing both exact-matching limitations and evolving user needs.

A key insight from the results is that adjusting the rating threshold significantly impacts the system's effectiveness. A lower threshold (3.0 out of 5) resulted in a near-perfect F1-score due to a drastic increase in recall, ensuring that more prompts were suggested to the user. However, this comes at the cost of precision, potentially introducing noise if users expect only high-confidence recommendations. In contrast, a higher threshold (3.5 out of 5) drastically reduced recall and F1-score, favouring precision but limiting the diversity of suggestions. This demonstrates that the optimal threshold depends on the intended balance between relevance and exploration. A practical implementation would likely require a tuneable threshold, allowing users to adjust recommendations based on their specific needs. Another challenge is ensuring that the system provides meaningful recommendations when an exact match is unavailable. Traditional recommendation systems struggle with user-generated prompts that deviate from predefined dataset entries. This is particularly problematic in a domain as nuanced as ML code generation, where users may describe similar concepts in different ways. Relying solely on exact matches restricts the system's ability to adapt to novel prompts, making it less useful for researchers working on cutting-edge topics.

To overcome this, we propose semantic similarity techniques that transform prompts into vector representations using models like BERT or GPT embeddings. This enables cosine similarity calculations to identify relevant prompts based on contextual meaning rather than exact wording. This approach significantly enhances the system's robustness, ensuring that even if a prompt is phrased differently, the user still receives useful recommendations. For example, a request for "privacy-preserving NLP techniques" could retrieve prompts related to differential privacy or federated learning, even if those terms were not explicitly mentioned. However, semantic similarity alone is not infallible. In cases where it fails to find a close match, fallback strategies are necessary. The system could default to providing the highest-rated prompts in the dataset or suggest commonly used ethical considerations. This ensures that users always receive some level of guidance, even when the input prompt is original. A hybrid approach combining semantic similarity with curated fallback recommendations would strike a balance between adaptability and reliability. A further enhancement to the system is dynamic dataset expansion. The ability to incorporate new prompts over time addresses a key limitation of static datasets, which may become outdated or fail to reflect the latest research trends. By allowing new user-generated prompts to be integrated and linked with inferred ratings, the recommender system evolves alongside advancements in ML and AI. This not only improves recommendation quality but also fosters a more inclusive and representative prompt database.

Beyond technical improvements, ethical considerations remain central to this recommender system. Traditional recommendation methods risk reinforcing biases present in historical data. By introducing filtering mechanisms that prioritise fairness and transparency, the system mitigates potential biases in prompt suggestions. Additionally, incorporating user feedback loops can refine recommendations, ensuring that prompts align with best practices in ethical AI development. The scalability of this approach is also noteworthy. As LLMs become more widely used for generating ML code, the demand for effective prompt recommendations will increase. A flexible recommender system capable of handling diverse user queries, evolving datasets, and ethical constraints ensures long-term usability. Future iterations could integrate real-time learning, where user interactions dynamically refine the model, improving recommendations over time.

In summary, this study presents a recommender system that enhances ethical prompt suggestions for ML code generation using LLMs. By balancing precision and recall, employing semantic similarity techniques, implementing fallback mechanisms, and enabling dataset expansion, the system remains adaptable and contextually relevant. These advancements ensure that users receive meaningful recommendations, even when dealing with complex, evolving topics. Furthermore, integrating ethical considerations into the recommendation process reduces bias and promotes responsible AI development. As AI research continues to progress, such systems could play an essential role in ensuring that ML model generation remains guided by ethical principles, fostering transparency, inclusivity, and accountability in AI-driven innovation.

In the future we plan to (a) incorporate explainable AI (XAI) algorithms because collaborative filtering effectively identifies related prompts, but users may struggle to understand the ethical reasoning behind each suggestion and (b) provide defences to enhance the robustness of the proposed system.

**Acknowledgement**

Funded by the European Union Horizon Europe programme CL2 2024-TRANSFORMATIONS-01-06 through ALFIE under Grant Agreement 101177912. Views and opinions expressed are however those of the author(s) only and do not necessarily reflect those of the European Union or the Agency. Neither the European Union nor the granting authority can be held responsible for them.